\documentclass{caosp309}

\usepackage{graphicx}

\usepackage{natbib}
\usepackage[flushleft]{threeparttable}
\usepackage{wrapfig}

\articleNo{XXX}
\pubyear{XXX}
\volume{XX}
\volnumber{x}
\firstpage{x}
\received{November 15, 2024}
\accepted{MM DD, YYYY}

\def\BibTeX{{\rm B\kern-.05em{\sc i\kern-.025em b}\kern-.08em
             T\kern-.1667em\lower.7ex\hbox{E}\kern-.125emX}}

\def\kms{km\,s$^{-1}$}

\begin{document}

%
\hauthor{M.\,Abdul-Masih}

\title{Observations of massive contact binaries in the local universe}


%
%
\author{
        M.\,Abdul-Masih\inst{1,2}\orcid{0000-0001-6566-7568}
       }

%
\institute{
    Instituto de Astrof\'isica de Canarias, C. V\'ia L\'actea, s/n, 38205 La Laguna, Santa Cruz de Tenerife, Spain\label{inst:iac}, \email{mabdul@iac.es}
    \and Universidad de La Laguna, Dpto. Astrof\'isica, Av.\ Astrof\'sico Francisco S\'anchez, 38206 La Laguna, Santa Cruz de Tenerife, Spain \label{inst:ull}
    }

\date{November 1, 2024}

\maketitle

\begin{abstract}
The contact phase represents a crossroad in the evolution of massive binary stars. Depending on the internal physics, the predicted end products can vary greatly including various exotic objects such as Be stars, magnetic massive stars, LBVs and gravitational wave sources. This phase also offers a unique observational laboratory to study binary interaction physics. Here, I review the current state of the field of massive contact binary observations.  I summarize the techniques available to identify and characterize these systems as well as the limitations of each and the potential biases that they introduce.  I present the sample of known confirmed systems and what the bulk statistics can tell us about their formation and evolution. Next I discuss the challenges that these systems pose from a characterization point of view and how we can overcome these. Finally I discuss the future direction of the field on the observational side.

\end{abstract}

%
\section{Introduction}\label{intr}

Binarity is common in massive stars and $\sim$70\% of all O-type stars are expected to interact with a companion at some point during their lifetime \citep{Sana2012, Moe2017}. Of the systems that interact, 40\% ($\sim$25\% of all O-type stars) will evolve through a contact phase and eventually merge \citep{Pols1994}. Population synthesis studies indicate that any given time, approximately 1\% of all O-type stars will be found in a contact configuration \citep{Henneco2024}, however, so far very few massive contact systems are known and even fewer have been studied in detail. 

The contact phase represents one of the most extreme forms of binary interactions throughout all branches of massive binary evolution. While in a contact configuration, the components share a common envelope through which mass and energy are exchanged and experience a variety of interaction processes including but not limited to tidal locking, mutual illumination, internal mixing, and angular momentum exchange \citep{Wellstein2001, deMink2007}. These binary interactions can fundamentally alter the evolutionary path of a system, affecting the main sequence lifetime, the interior structures of the components, and the positions on the Hertzsprung Russell diagram \citep{Marchant2024}. That said, depending on the treatment of these internal processes and the interaction physics, the theoretically predicted end products of systems that evolve through a contact phase vary widely, including objects such as magnetic massive stars \citep{Schneider2019, hubrig2024}, Be stars \citep{Shao2014}, luminous blue variables \citep{Smith2018}, blue stragglers \citep{Mateo1990} and double black hole binaries \citep{Marchant2016, Mandel2016, deMink2016}. Without stringent observational constraints during the contact phase, it is impossible to know which of these evolutionary channels will dominate. Over the past few years, there has been a renewed effort to better understand these systems from both the theoretical and observational side.  Here, I focus on outlining the current state of the field on the observational side.

\section{Detection and characterization}

\begin{figure}
\centerline{\includegraphics[width=1.0\textwidth,clip=]{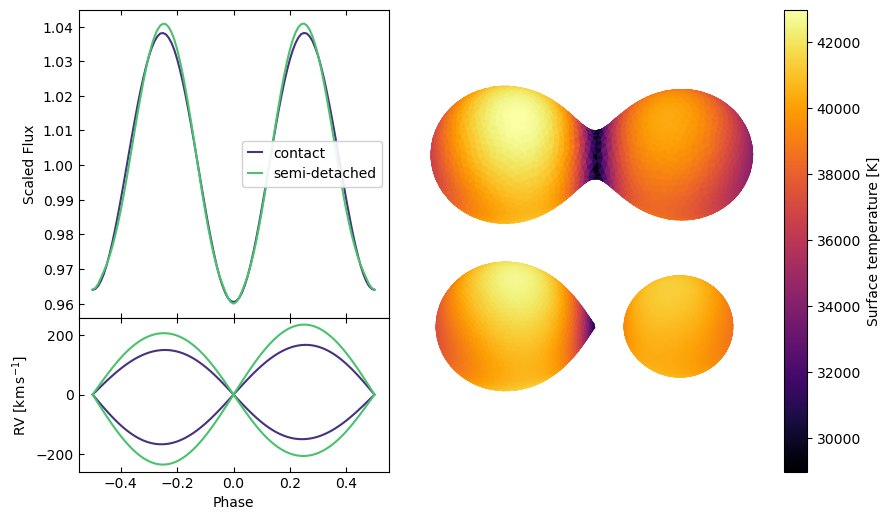}}
\caption{Left: simulated light curve and radial velocity curves for a contact binary (blue) and semi-detached binary (green) using PHOEBE. Right: morphologies of each system simulated on the left panel are displayed with the color representing the surface temperature. Both systems have the same period, semi-major axis, mass ratio and primary temperature, however the secondary temperature, inclination and stellar radii differ.}
\label{fig:lc_morph}
\end{figure}

Massive contact binaries are often identified by their short periods ($\sim$ 0.5-3 days) and large amplitude sinusoidal-like light curves.  Depending on the mass ratio, fillout factor \citep[see ][]{Mochnacki1972}, temperature ratio and inclination, the exact morphology of the light curve can change, but in general contact binaries have smooth light curves without sharp features (except in the case of a total eclipse).  The primary and secondary eclipse depths are usually fairly similar (but not necessarily equal) and do not show as strong of variations as other classes of binaries do. Unfortunately, however several other types of objects show similar light curve morphologies making the identification and confirmation of contact systems difficult.  Fig. \ref{fig:lc_morph} shows the morphology of the light curve and radial velocity curves for a contact system and a semi-detached system generated using the PHOEBE code \citep{Prsa2016, Horvat2018, Conroy2020, Jones2020}.  In this case, the light curves show almost identical morphologies despite the fact that the binary configurations are quite different. Looking at the radial velocity curves, however, there is a much more significant difference between the semi-amplitudes of each component. Similar comparisons can be made between contact systems and detached ellipsoidal variable systems, however, again in this case, the radial velocity curve can allow one to distinguish between these configurations

The most sure fire way to confirm that a given system is in contact is to directly resolve the stellar surface. This could be done with interferometry, however given the expected angular sizes and magnitudes of the currently known massive contact binary candidates, this is not yet possible. The next best method is performing a simultaneous light curve and radial velocity curve (LC+RV) fit using an appropriate eclipsing binary modelling code.  Several such codes exist, however only the Wilson-Devinney code \citep{Wilson1971, Wilson1979, Wilson2008, Wilson2014} and the PHOEBE code \citep{Prsa2005, Prsa2016, Horvat2018, Conroy2020, Jones2020} are able to model all binary morphologies ranging from well-detached up to and including contact systems. These different morphologies each require different set-ups within the codes themselves, so, confirming the configuration of a contact binary requires one to model both the light curve and the radial velocity curve assuming a detached, a semi-detached and a contact configuration in order to determine which results in the best fit.  For the purposes of this paper, I will consider a system to be a confirmed massive contact binary only if it has been characterized via a combined LC+RV fit.  Systems that have been characterized via any other means will be referred to as candidate systems.  For the remainder of the paper, unless explicitly stated otherwise, I will be discussing only the confirmed systems.

\section{Observational overview}

So far only 13 massive contact systems have been confirmed via a combined LC+RV fit in the literature and these are listed in Tab. \ref{tab:contacts}.  It should be noted that there are additional objects with LC+RV fits for contact configurations in the literature, however these have later been discarded as contact systems for various reasons and are therefore not included in the sample in Tab. \ref{tab:contacts}.  

\begin{table}[t] 
\begin{threeparttable}
\small
\begin{center}
\caption{Confirmed massive contact binaries with full LC+RV fits.}
\label{t1}
\begin{tabular}{lcrccccl}
\hline\hline
Name  & Sp. type & $\mathrm{V}_\mathrm{mag}$ & Period & $q$ & $f$ & Multiplicity & Ref.\\
  &  &   & [days] &  & &  & \\
\hline
HD 64315 & O+O & 9.2 & 1.019 & 1.00 & 1.31 & 4 & 1 \\
LSS 3074 & O+O & 11.7 & 2.1852 & 0.86 & 1.05 & 2 & 2 \\
MY Cam & O+O & 9.8 & 1.1755 & 0.84 & 1.01 & 2 & 3 \\
MY Ser & O+O & 7.5 & 3.3216 & 0.95 & 1.99 & 3 & 4,5,6 \\
SMC 108086$^\dagger$ & O+O & 14.9 & 0.8831 & 0.85 & 1.7 & 2 & 7 \\
TU Mus & O+O & 9.3 & 1.3873 & 0.62 & 1.12 & 3 & 8,9 \\
V382 Cyg & O+O & 8.7 & 1.8855 & 0.73 & 1.1 & 3 & 9,10,11 \\
V729 Cyg & O+O & 9.2 & 6.598 & 0.29 & 1.22 & 6 & 6,12,13 \\
VFTS 352 & O+O & 14.5 & 1.1241 & 0.99 & 1.29 & 2 & 14 \\
CT Tau & B+B & 10.3 & 0.6668 & 0.98 & 1.99 & 3 & 15 \\
GU Mon & B+B & 11.6 & 0.8966 & 0.98 & 1.72 & 3 & 15 \\
V701 Sco & B+B & 9.0 & 0.7619 & 1.00 & 1.55 & 3 & 15 \\
V745 Cas & B+B & 8.1 & 1.4106 & 0.57 & -- & 5 & 16 \\
\hline\hline
\end{tabular}
\begin{tablenotes}
    \small {
    \item{\textbf{Notes:} $\dagger$ full name: OGLE SMC-SC10 108086}
    \item \textbf{References:} (1) \citet{Lorenzo2017}, (2) \citet{Raucq2017}, (3) \citet{Lorenzo2014}, (4) \citet{Ibanoglu2013}, (5) \citet{Leitherer1987}, (6)\citet{Lanthermann2023}, (7) \citet{Hilditch2005}, (8) \citet{Penny2008}, (9) \citet{Qian2007}, (10)\citet{Harries1997}, (11) \citet{Martins2017}, (12) \citet{Yasarsoy2014}, (13) \citet{Rauw2019}, (14) \citet{Almeida2015}, (15) \citet{Yang2019}, (16) \citet{cakirli2014}
    
            }
\end{tablenotes}
\label{tab:contacts}
\end{center}
\end{threeparttable}
\end{table}

\subsection{Sample statistics}

\begin{figure}
\centerline{\includegraphics[width=1.0\textwidth,clip=]{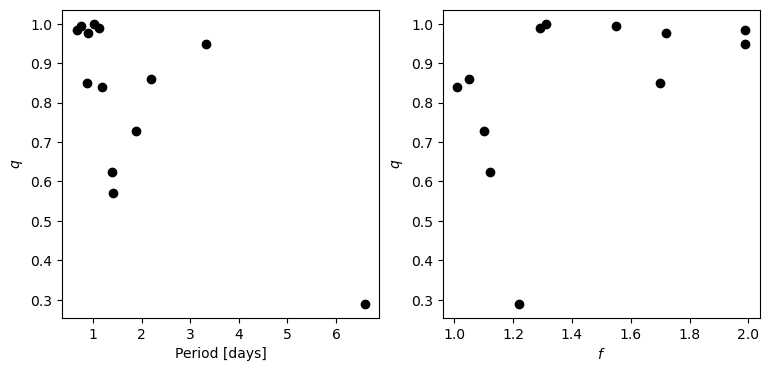}}
\caption{Mass ratio versus orbital period (left) and mass ratio versus fillout factor (right) for the confirmed massive contact systems.}
\label{fig:pqf}
\end{figure}

Based on Tab. \ref{tab:contacts}, we can already start to look for trends in the sample to see how these compare with theoretical predictions for this phase.  Looking at the sample, all but one object have periods ranging from 0.8-3 days with most concentrating towards the lower end of the period range, however a distinct outlier can be seen in V729 Cyg, which has a period of more than 6 days.  This system is classified as a blue supergiant \citep{Sota2011} and is therefore more evolved than the rest of the sample so it is likely that this system did not originate from Case A mass transfer as is the case for the rest of the systems.  That said, binary evolution studies predict that there should be more Case B and Case C contact systems formed than Case A contact systems \citep{Henneco2024}, however the shorter lifetimes of these systems may explain why so many more Case A systems are currently known.  This could also be due to a selection bias as contact systems with more evolved components are expected to have longer periods, and therefore may not fall within the selection criteria used when searching for contact systems. 

The other two important parameters that we can compare with theoretical predictions are the mass ratio and the fillout factor.  Both \citet{Marchant2016} and \citet{Menon2021} calculated detailed evolutionary models and performed population synthesis studies focused on contact systems, and both predict that after coming into contact, systems should quickly evolve (on a thermal time scale) towards a mass ratio of unity and then continue to evolve on the nuclear timescale.  This would suggest that we would expect to see orders of magnitude more objects with $q$=1, however observationally, this does not seem to be the case.  As is shown in Fig. \ref{fig:pqf}, the mass ratios are fairly well distributed between 0.3 and 1, and while there is certainly an overdensity at $q$=1, it is not nearly as extreme as theoretical studies predict. Based on the right panel of Fig. \ref{fig:pqf}, which shows the mass ratio as a function of the fillout factor (which can be used as an evolutionary proxy as the fillout factor will increase as the components evolve and expand), there does appear to be a trend towards a mass ratio of 1 as the fillout factor increases, however the scatter is still fairly large.  Studies of the rate of period change can provide useful insights in this case as the theoretical predictions state that systems with mass ratios significantly far from 1 should be transferring mass at a higher rate, which in turn will affect the period accordingly.  Observational studies seem to indicate that high mass contact systems show very stable periods that are evolving on the nuclear time scale independent of the mass ratio \citep{Yasarsoy2014, Li2022, Abdul-Masih2022, Vrancken2024}.  This combined with the right panel of Fig. \ref{fig:pqf} may indicate that either massive contact systems are evolving towards a mass ratio of unity but at a much slower rate than what is expected, or that they may never reach a mass ratio of 1 before merging.

\subsection{Stellar properties}
Of the 13 confirmed systems, only 4 have been studied in detail spectroscopically to obtain atmosphere parameters and abundances \citep{Raucq2017, Martins2017, Abdul-Masih2019, Abdul-Masih2021}. A striking result from these works is that in all cases, the abundances of the CNO elements are consistent with the baseline abundances of the regions where the objects are found, indicating little to no CNO processed material on the surface of the stars. Given the extreme rotation rates of these systems ($\ge$ 300\kms) this is surprising as rotational mixing is expected to efficiently mix core processed material to the surface at these rotation rates \citep{Maeder2000a}. The observations also indicate that in general both the primary and secondary components are overluminous and hotter than expected given their masses, which is usually interpreted as a sign of efficient internal mixing. These two observational findings appear to be at odds with one another, and it is still unclear how these fit together \citep[see discussion in ][]{Abdul-Masih2021}.

Another important finding from these studies is that the secondary component is always found very close to the primary on the Hertzsprung-Russell diagram, independent of the mass ratio of the system \citep{Raucq2017, Abdul-Masih2021}. This effect is much too strong to be explained by reflection effects, indicating that there must be a mechanism that efficiently transports energy through the common envelope between the two components. Similar effects have been observed in low mass contact systems and are attributed to sideways convection \citep{Lucy1968a, Lucy1968b}, however in the case of massive stars the envelopes are radiative so the mechanism is certainly different.  Theoretical studies including heat transfer through the radiative envelope have shown that this effect can affect the distribution of mass ratios, however this effect alone is still not enough to rectify the discrepancy between the observed sample and the theoretical predictions \citep{Fabry2022, Fabry2023, Fabry2024}.

\subsection{Multiplicity}

While no dedicated multiplicity study focused on massive contact systems has been conducted thus far, many confirmed and candidate systems have been discovered to be in higher order multiple systems.  The majority of these detections come through eclipse timing variations (ETVs) due to the light travel time effect \citep[e.g.,][]{Qian2007, Zasche2017, Yang2019, Li2022}, however some have been detected as spectroscopic multiples \citep[e.g.,][]{Lorenzo2017, Janssens2021} or through interferometry \citep[e.g., ][]{Lanthermann2023}. Systems with a companion that has been discovered via ETVs show outer orbital periods of between $\sim$ 5-80 years, however companions discovered through spectroscopy and interferometry do not have well constrained orbital periods as of yet.  Of the confirmed contact systems, almost 70\% of the sample have at least one confirmed companion and about 25\% have more than one additional companion. That said, the 4 systems that currently do not have a detected companion have been discovered within the last 10 years and lack sufficient archival data to perform an ETV study, and given the periods of the outer companions discovered with this method, a much longer baseline may be needed to detect any companions.  Therefore, this 30\% of contact systems without additional companions should be viewed as an upper limit as additional data and time will allow us to probe the parameter space where the companions should be found if they exist.

\begin{wrapfigure}{R}{0.5\textwidth}
  \vspace{-15pt}
  \centerline{\includegraphics[width=0.48\textwidth]{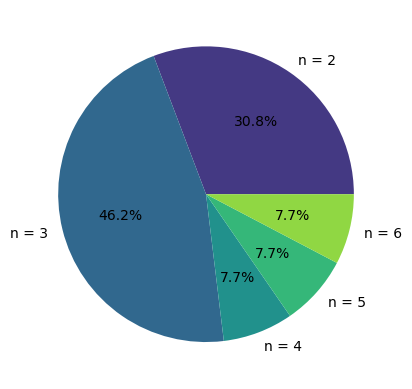}}
  \caption{Multiplicity statistics for confirmed massive contact sample}
  \vspace{-15pt}
\end{wrapfigure}

Comparing the multiplicity statistics of the confirmed massive contact sample with those of \citet{Offner2023}, the contacts triple fraction is already rather high considering that it contains both B-type and O-type systems.  Considering that this is a lower limit on the triple fraction, it is possible that the multiplicity properties of contact systems are fundamentally different from that of other O-type systems, however this is not entirely unexpected due to the compactness of the inner orbit. Given the orbits of the companions, its possible that the companion may play a role in the formation of the contact system, but a deeper investigation is needed to determine the potential evolutionary effects that the companion causes.


\section{Challenges}

The analysis techniques and tools that we use to characterize contact systems each contain their own set of assumptions and limitations, and it is important to understand how these can affect our understanding of the systems that we apply them to. This applies first and foremost to the techniques that we use to determine the binary configuration for these systems.  For example, while the Wilson-Devinney and PHOEBE 1.0 codes are both able to model contact systems, these codes utilize a trapezoidal meshing that can often lead to gaps in the surface coverage at the bridge. The triangluated meshing used by the PHOEBE 2 code avoids this issue and is able to completely cover the surface, making it a better option for modelling contact systems \citep{Prsa2016}.  Another important aspect of binary modelling that is often overlooked in codes is the gravity darkening prescription.  All of the currently available eclipsing binary modeling codes use the von Zeipel gravity darkening prescription \citep{vonZeipel1924} to relate the local surface temperature to the local surface gravity, which while sufficient for well detached binaries, leads to very important systematic biases when applied to highly deformed systems.  Improved gravity darkening prescriptions have been developed for rapidly rotating stars and (semi-)detached binaries \citep{EspinosaLara2011, EspinosaLara2012}, however so far no such prescription has been developed for contact systems.  The von Zeipel gravity darkening prescription leads to more extreme temperatures than is realistic meaning that less deformed systems are needed to reproduce the light curves.  By implementing new gravity darkening prescriptions, it is possible that many systems that are currently thought to be semi-detached or ellipsoidal variables may end up becoming contact systems.  Until updated gravity darkening prescriptions are developed, however, other clues may help us to determine whether an edge case is actually in contact or not.  \citet{Vrancken2024} showed that the period stability may be used to differentiate between contact systems and semi-detached systems which are expected to have higher rates of period change.  Additionally, similar component temperatures in unequal mass systems may indicate that a system is in contact.

The spectroscopic analysis techniques that have traditionally been applied to these types of systems also have several important assumptions that are not necessarily valid in the case of contact systems.  Usually massive close binaries are analyzed by first disentangling the spectra to obtain the spectral contribution of each component and then fitting each separately using an appropriate radiative transfer code.  Spectral disentangling assumes that the morphology of the spectra remains constant over the orbit, and that the only things that vary are the radial velocity and light ratio.  This in essence enforces a spherical assumption on the system and does not account for the temperature variations across the surface.  This spherical assumption is then reinforced during the fitting process as the available NLTE radiative transfer codes are almost exclusively one-dimensional.  To combat this, new spectroscopic techniques have been developed that patch 1D stellar atmospheres onto a 3D surface, which approximates how the different surface regions contribute to the combined integrated spectrum of the system \citep[see e.g., ][]{Palate2012, Abdul-Masih2020a}. Unfortunately, these codes still suffer from the gravity darkening problem discussed above, but overall these codes are more robust and suffer from fewer assumptions and biases than the disentangling methods do.

As demonstrated in \citet{Abdul-Masih2021, Abdul-Masih2023}, accounting for the surface geometry can have an important effect on the derived surface parameters.  When comparing the temperatures and abundances obtained using the more traditional disentangling method with those obtained using the spectroscopic patch model SPAMMS \citep{Abdul-Masih2020a}, major discrepancies were found that appeared to be dependent on the inclination of the system.  This was later confirmed in \citet{Abdul-Masih2023}, which focused on single rapidly rotating stars.  Here, it was shown that the temperature measured using 1D models could vary by up to 3000K depending on the inclination of the system.  Furthermore, failing to account for the surface geometry also heavily biased the abundances, leading to a systematic offset in the helium abundance by more than a factor of 2 in some cases.  These studies demonstrate the importance of not only determining the geometry of the system accurately, but accounting for this geometry when measuring atmospheric parameters and surface abundances.

\section{Conclusions and future prospects}

While the number of confirmed massive contact binaries is far below theoretical expectations, many interesting results have come from these 13 sources already.  The sample is well distributed in period, mass ratio and fillout factor, and even though the sample size is rather small, the distribution is already challenging theoretical predictions.  The fillout factors, which can be used as a proxy for the time since the initiation of the contact phase, show a loose correlation with the mass ratios suggesting that systems may tend to mass ratios of 1, but on significantly longer time scales than what the theoretical models predict.  This is further supported by period stability measurements which show that the systems appear to be stable on the nuclear time scale. Results of spectroscopic analysis also seem to contradict theoretical predictions, as all systems studied so far seem to have baseline abundances, indicating little to no internal mixing.  The temperature and luminosities, however, are higher than expected, indicating efficient mixing, presenting an interesting contradiction that has still not been rectified. The sample also shows a high triple fraction of about 70\%, however, it is likely that the triple fraction is even higher considering that many systems have been discovered and characterized recently.  It is unclear if these companions have affected the past evolution of these systems, so more dedicated evolutionary modelling is needed to better understand the role of third bodies.

While much progress has been made, there are still many open questions and many avenues for continued investigation.  First and foremost, expanding the sample and obtaining a realistic occurrence rate of this phase is of tantamount importance.  This includes both searching for new systems and characterizing the systems that have yet to be confirmed. On this front, implementing new gravity darkening prescriptions into our eclipsing binary modelling codes could change the picture somewhat, as this is still one of the largest sources of uncertainty when it comes to determining the configuration of a given system.  For the systems that have already been confirmed, spectroscopic analysis of the sample may also reveal useful insights into the evolutionary histories of the systems.  Only 4 of the 13 confirmed massive contact systems have been studied in detail spectroscopically, so it is important to analyze the rest of the sample to determine whether the observed abundance and temperature trends hold for the entire sample. This will allow us to better understand when these systems came into contact and whether or not the initial mass transfer leading up to the contact configuration was conservative. On a similar front, characterizing the triple companions around contact systems could also prove to be important to understand their past or future evolution. Overall, we are entering an exciting time in the field of massive contact binaries.  We are on the cusp of being able to analyze our sample in a statistical way, which will could lead to some exciting discoveries in the next few years.

\acknowledgements
This project received the support from the ``La Caixa'' Foundation (ID 100010434) under the fellowship code LCF/BQ/PI23/11970035.

\bibliography{mybib}

\end{document}